# Blind CT Image Quality Assessment Using DDPM-derived Content and Transformer-based Evaluator

Yongyi Shi, Wenjun Xia, Ge Wang, Xuanqin Mou

*Abstract* — **Lowering radiation dose per view and utilizing sparse views per scan are two common CT scan modes, albeit often leading to distorted images characterized by noise and streak artifacts. Blind image quality assessment (BIQA) strives to evaluate perceptual quality in alignment with what radiologists perceive, which plays an important role in advancing low-dose CT reconstruction techniques. An intriguing direction involves developing BIQA methods that mimic the operational characteristic of the human visual system (HVS). The internal generative mechanism (IGM) theory reveals that the HVS actively deduces primary content to enhance comprehension. In this study, we introduce an innovative BIQA metric that emulates the active inference process of IGM. Initially, an active inference module, implemented as a denoising diffusion probabilistic model (DDPM), is constructed to anticipate the primary content. Then, the dissimilarity map is derived by assessing the interrelation between the distorted image and its primary content. Subsequently, the distorted image and dissimilarity map are combined into a multi-channel image, which is inputted into a transformer-based image quality evaluator. Remarkably, by exclusively utilizing this transformer-based quality evaluator, we won the second place in the MICCAI 2023 low-dose computed tomography perceptual image quality assessment grand challenge. Leveraging the DDPM-derived primary content, our approach further improves the performance on the challenge dataset.**

## I. INTRODUCTION

X-RAY image quality assessment (IQA) plays a crucial role in computed tomography (CT) imaging, facilitating the advancement of novel algorithms for low-dose CT reconstruction. Two strategies for low-dose CT scans are either reducing the X-ray tube current or acquiring sparse views. However, these strategies often introduce noise and/or streak artifacts in the filtered backprojection (FBP) images. To address these problems, deep learning-based methods have been used for low-dose CT image denoising [1-5] and sparse view CT reconstruction [6-10]. When the quality of these reconstructed images is assessed, radiologist's opinions serve as the gold standard. Nonetheless, collecting these opinions is an expensive and intricate process, making it impractical for real-time and large-scale IQA tasks. Consequently, peak signal-to-noise ratio (PSNR) and structural similarity index measure (SSIM) have been widely used as surrogate metrics [11]. However, both PSNR and SSIM have shown limited correlation with radiologists' opinions on image quality, primarily due to their reliance on mathematical models that do not account for the intricacies of human perception. Moreover, the requirement for reference images to calculate these metrics poses challenges in clinical environments; for example, obtaining high-quality images is infeasible without increasing patient radiation exposure. To overcome these limitations, one direction is to develop no-reference image quality metrics that correlates well with radiologists' opinion on image quality.

No-reference image quality assessment (NR-IQA), also known as blind IQA (BIQA), is widely used to evaluate the quality of natural images [12-13]. Traditional BIQA methods typically comprise three steps. First, some handcrafted descriptors are employed to extract quality-aware features of training images. Then, the statistical distribution of the extracted features serves as the guidance. Some approaches also involve parameterizing this distribution through modeling. Finally, a mapping function, such as support vector regression (SVR) [14], is designed to convert the distributions into a quality score. In traditional BIQA, features are derived from various sources including discrete wavelet transform (DWT) [15-17], discrete cosine transform (DCT) [18] and spatial domain measures [19-20] to predict the perceptual quality. Alternatively, it is assumed that the human visual system (HVS) gauges image quality by discerning features like gradient [21], luminance contrast [22] or local binary pattern [23]. Certain approaches combine these extracted features to improve image quality assessment further [24-25]. Nevertheless, due to the complexities of image contents and distortion patterns, the representation capabilities of handcrafted features are often unsatisfactory.

In recent years, the convolutional neural network (CNN) has gained major attention in BIQA tasks because of its potent

This work was supported in part by the National Institute of Biomedical Imaging and Bioengineering/National Institutes of Health under Grant R01 EB016977. The work of Xuanqin Mou was supported in part by the Natural Science Foundation of China, under Grant 62071375. (Corresponding authors: Ge Wang; Xuanqin Mou).

Yongyi Shi, Wenjun Xia and Ge Wang are with the Department of Biomedical Engineering, Rensselaer Polytechnic Institute, Troy, NY 12180 USA (e-mail: shiy11@rpi.edu; xiaw4@rpi.edu; wangg6@rpi.edu).

Xuanqin Mou are with the Institute of Image Processing and Pattern Recognition, Xi'an Jiaotong University, Xi'an 710049, China (e-mail: xqmou@mail.xjtu.edu.cn).

feature representation power [26-27] and started gaining momentum [28-33]. Rank-IQA utilizes synthetically generated distortions and a Siamese network to rank images based on their quality [29]. DBCNN demonstrates effectiveness with both synthetic and authentic image distortions [30]. Hyper-IQA segregates features into low-level and high-level categories, then transforms the latter to reshape the former's influence [31]. Meta-IQA employs meta-learning to train the networks on distinct types of distortions, thereby acquiring prior knowledge [32]. More recently, vision transformers (ViTs) [33] have emerged as competitive alternatives to CNNs. Their self-attention mechanism empowers ViT to grasp global contextual information from an entire image, ensuring a comprehensive consideration of all image features of significance for task-specific prediction. This attribute aligns well with the requirements of BIQA task, which involves predicting a quality index globally [34-40]. MSTIQA leverages a Swin transformer to amalgamate features from multiple stages to enhance quality assessment [38]. MANIQA introduces a multi-dimensional attention network for BIQA. This approach introduces the transposed attention block and the multiscale Swin transformer block to strengthen global and local interactions [39]. MAMIQA employs a lightweight attention mechanism that utilizes decomposed large kernel convolutions to extract multiscale features. Furthermore, it includes a feature enhancement module to enrich local fine-grained details and global semantic information at multi-scales [40]. Among the various transformer-based BIQA methods, MANIQA has not only achieved the state-of-the-art performance in BIQA tasks but also secured the first place in the no-reference track of NTIRE 2022 perceptual image quality assessment challenge [41]. Intriguingly, the top three methods in this challenge all rely on transformer-based techniques, highlighting the efficacy of the transformers in BIQA tasks. Nonetheless, the lack of reference information poses a challenge, preventing these methods from aligning seamlessly with the HVS and may adversely impacting their overall performance.

The theory of internal generative mechanisms (IGMs) suggests that the HVS engages in an active process of deducing the primary content of an image during human evaluations [42-44]. In this process, IGM initially analyzes pixel correlations within an input image. In conjunction with intrinsic prior knowledge, IGM deduces the corresponding primary content as an active comprehension of the input image [45]. The primary content consists of essential scene information, representing the structured, meaningful elements within the image, which is transported to the high level of HVS for interpretation [46]. Thus, the ability to generate high-quality primary content becomes critical in emulating the HVS. Notably, recent successes in adapting and applying the denoising diffusion probabilistic model (DDPM) and other diffusion models have showcased their amazing capabilities in both image generation and restoration [47- 49]. These qualities underscore a huge potential of diffusion models in generating high-quality primary content, particularly in the context of low-dose CT imaging [50, 51] which is featured by uniquely intricate noise and streak artifacts, quite different from what in natural images.

Low-dose CT images frequently suffer from conspicuous noise and streak artifacts, being the primary culprits of the image quality deterioration. Specifically, the streak artifacts exhibit directional appearance contributing to globalized artifacts in a whole image [6]. This makes the existing datasets of natural images inadequate for accurately predicting quality scores of low-dose CT images. On the other hand, given the scarcity of open datasets for low-dose CT BIQA, experiments have been conducted using disparate datasets, yielding results that are challenging to compare and interpret [52, 53]. These challenges demand a standardized image quality metric in general, and for low-dose CT imaging in particular. Notably, a low-dose CT BIQA dataset has recently been unveiled for the MICCAI 2023 low-dose CT perceptual image quality assessment grand challenge [54]. As low-dose CT images are obtained by reducing the number of projections per rotation and/or the X-ray tube current, the combination of sparse view streaks and noise needs to be dealt with in the challenge so that the best-performing IQA model can be identified and made applicable in real clinical environments.

Given the globalized artifact patterns in low-dose CT images, we utilized a transformer-based quality evaluator [39] and won the second place in the MICCAI 2023 low-dose CT perceptual image quality assessment grand challenge [54]. We found in our study that directional artifacts may sometimes be misleading by resembling genuine anatomical structures. To further improve the performance, inspired by IGM here we propose a novel approach utilizing a DDPM-based active inference for low-dose CT BIQA. At the outset, given DDPM's capability to simultaneously eliminate noise and streak artifacts, we employ DDPM to generate high-quality primary content, closely mimicking the HVS. Recognizing the heightened sensitivity of HVS to structures, the dissimilarity map is extracted from the distorted image and its primary content. After that, incorporating such diverse prior information as input, a multi-channel image is synthesized, amalgamating content, distortion, and structural characteristics for comprehensive quality prediction. The final step involves employing a transformer-based quality evaluator to predict the score. Empirical trials conducted on the low-dose CT BIQA challenge dataset systematically affirm the efficacy of our proposed method.

In brief, the main contributions of this paper can be summarized as follows:

*1)* We propose a conditional DDPM to emulate the active inference process of IGM. The proposed DDPM-based active inference module can effectively predict the primary content of distorted images which contain intricate noise and streak artifacts from both low-dose and sparse view CT imaging.

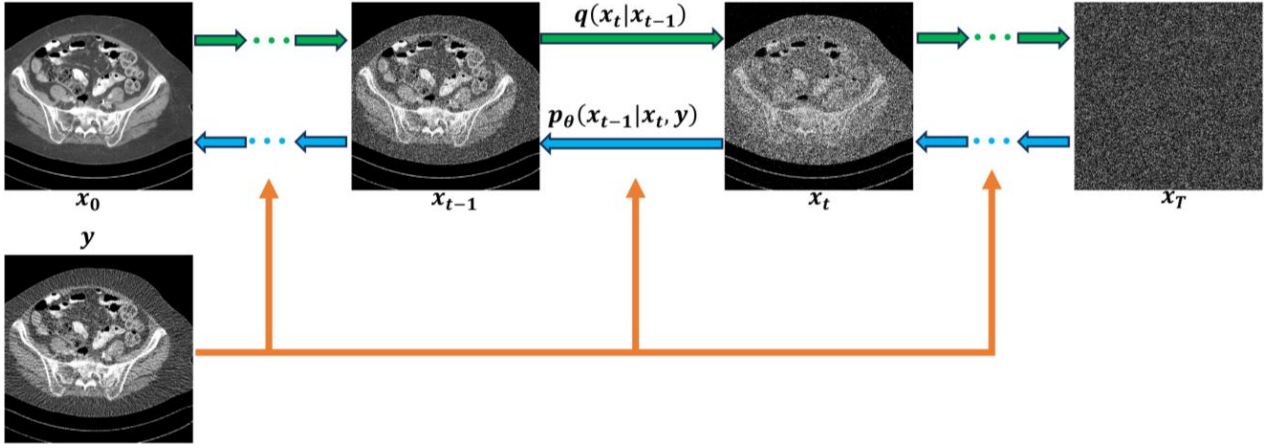

Figure 1. Workflow of the conditional DDPM that maps low-dose CT images to normal dose counterparts.

2) Based on the primary content, we introduce a transformer-based quality evaluator to predict the image quality on a low-dose CT BIQA dataset. Note that we secured the runner-up position in the MICCAI 2023 low-dose CT perceptual image quality assessment grand challenge by only employing this transformer-based quality evaluator. On that basis, we have improved the image quality assessment performance even further, as explained in detail below.

The remainder of the paper is organized as follows. Section II outlines the process of utilizing a conditional DDPM to acquire both the primary content and the dissimilarity map, while also detailing the transformer-based quality evaluator. Section III reports our evaluation results on the low-dose CT BIQA challenge dataset. Section IV discusses relevant issues and makes the conclusion.

## II. METHODOLOGY

### A. Conditional DDPM for Low-Dose CT

To generate high-quality primary content for emulating the HVS, a typical deep learning method trains a network to learn a mapping from low-dose CT images to normal-dose CT images. Let us assume that $y \in \mathbb{R}^N$ and $x \in \mathbb{R}^N$ are paired low-dose CT and normal-dose CT images. The parameters of the network can be trained as follows:

$$\min_\theta \|D_\theta(y) - x\|_2^2 \qquad (1)$$

where $D_\theta$ is the network parameterized by $\theta$, which is the key for producing high-quality images.

Recently, DDPM has demonstrated superior performance in generating high-quality images from their distorted counterparts, particularly in complex scenarios involving multiple distortions, such as CT imaging.

The architecture of the conditional DDPM is shown in Fig. 1. DDPM starts with a forward process that gradually adds noise to the normal-dose CT image $x_0 \sim q(x_0)$ over the course of $T$ timesteps according to a variance schedule $\beta_1, \cdots, \beta_T$:

$$q(x_t|x_{t-1}) = \mathcal{N}(x_t; \sqrt{1-\beta_t}x_{t-1}, \beta_t I) \qquad (2)$$

$$q(x_{1:T}|x_0) = \prod_{t=1}^{T} q(x_t|x_{t-1}) \qquad (3)$$

where $x_1, \cdots, x_T$ are latent variables of the same dimensionality as the sample $x_0 \sim q(x_0)$.

According to the properties of the Gaussian distribution, the sampling result $x_t$ at an arbitrary timestep $t$ can be written in the following closed form:

$$q(x_t|x_0) = \mathcal{N}(x_t; \sqrt{\bar{\alpha}_t}x_0, (1-\bar{\alpha}_t)I) \qquad (4)$$

where $\alpha_t = 1 - \beta_t$ and $\bar{\alpha}_t = \prod_{i=1}^{t} \alpha_i$.

After the forward process, $x_T$ follows a standard normal distribution when $T$ is large enough. Thus, if we know the conditional distribution $q(x_{t-1}|x_t)$, we can use the reverse process to get a sample under $q(x_0)$ from $x_T \sim \mathcal{N}(0, I)$. However, $q(x_{t-1}|x_t)$ depends on the entire data distribution, which is hard to calculate. Hence, a neural network was designed to learn a latent data distribution by gradually denoising a normal distribution variable, which corresponds to learning the reverse process of a fixed Markov Chain of length $T$ conditioned on a low-dose CT image $y$. The reverse process can be defined as:

$$p_\theta(x_{t-1}|x_t, y) = \mathcal{N}(x_{t-1}; \mu_\theta(x_t, y, t), \sigma_t^2 I) \qquad (5)$$

$$p_\theta(x_{0:T}|y) = p(x_T) \prod_{t=1}^{T} p_\theta(x_{t-1}|x_t, y) \qquad (6)$$

where $p(x_T)$ is the density function of $x_T$. In Eq. (6), $\mu_\theta(x_t, y, t)$ and $\sigma_t^2$ are needed to solve $p_\theta(x_{t-1}|x_t, y)$. According to the Bayes theorem, the posterior $q(x_{t-1}|x_t, x_0)$ are defined as

$$q(x_{t-1}|x_t, x_0) = \mathcal{N}(x_{t-1}; \tilde{\mu}_t(x_t, x_0), \sigma_t^2 I) \qquad (7)$$

where

$$\tilde{\mu}_t(x_t, x_0) = \frac{\sqrt{\alpha_t}(1-\bar{\alpha}_{t-1})}{1-\bar{\alpha}_t}x_t + \frac{\sqrt{\bar{\alpha}_{t-1}}(1-\alpha_t)}{1-\bar{\alpha}_t}x_0 \qquad (8)$$

$$\sigma_t^2 = \frac{(1-\bar{\alpha}_{t-1})(1-\alpha_t)}{1-\bar{\alpha}_t} \quad (9)$$

Since $\sigma_t^2$ is a constant, the most natural parameterization of $\mu_\theta(x_t, y, t)$ is a neural network that predicts $\tilde{\mu}_t(x_t, x_0)$ directly. Alternatively, given that $x_t = \sqrt{\bar{\alpha}_t}x_0 + \sqrt{1-\bar{\alpha}_t}\epsilon$, $\epsilon \sim \mathcal{N}(0, I)$, the posterior expectation in Eq. (8) can be expressed as

$$\tilde{\mu}_t(x_t, x_0) = \tilde{\mu}_t\left(x_t, \frac{1}{\sqrt{\bar{\alpha}_t}}(x_t - \sqrt{1-\bar{\alpha}_t}\epsilon)\right)$$

$$= \frac{1}{\sqrt{\alpha_t}}\left(x_t - \frac{1-\alpha_t}{\sqrt{1-\bar{\alpha}_t}}\epsilon\right) \quad (10)$$

Since $\tilde{\mu}_t(x_t, x_0)$ can be represented by $\epsilon$, we can also use a neural network model $D_\theta$ to predict the noise $\epsilon$, which has been proved to work well by Ho et al. [47]. Hence, the corresponding objective can be simplified to

$$\mathcal{L} = \mathbb{E}_{x,y}\mathbb{E}_{\epsilon,t}\left[\frac{(1-\alpha_t)^2}{2\sigma_t^2\alpha(1-\bar{\alpha})}\left\|\epsilon - D_\theta(\sqrt{\bar{\alpha}_t}x_0 + \sqrt{1-\bar{\alpha}_t}\epsilon, y, t)\right\|_2^2\right] \quad (11)$$

with $t$ uniformly sampled as $\{1,\cdots,T\}$.

In this study, the latent space is diffused into Gaussian noise using $T = 1000$ steps. A U-Net [55] model $D_\theta$ is trained to predict the noise $\epsilon$ in the latent space. To obtain high-quality primary contents, samples can be computed as follows:

$$x_{t-1} = \frac{1}{\sqrt{\alpha_t}}\left(x_t - \frac{1-\alpha_t}{\sqrt{1-\bar{\alpha}_t}}D_\theta(x_t, y, t)\right) + \sigma_t z \quad (12)$$

where $z \sim \mathcal{N}(0, I)$. For clarify, the pseudo codes of Algorithms 1 and 2 are presented for training and inference, respectively.

Through the emulation of IGM for active inference using conditional DDPM, we successfully obtained the primary content. It is important to note that conditional DDPM is not intended to completely restore a distorted image to the reference image, but the predicted primary content may present the reference image much better than the distorted input. Therefore, our goal is to estimate a dissimilarity map from this primary content, rather than merely quantifying the disparity between the distorted image and its primary content. The process for generating this dissimilarity map is depicted in Fig. 2.

After we obtain the primary content with the conditional DDPM, we calculate the SSIM map between the distorted image and the primary content, which assigns values between 0 and 1 to each and every pixel, with 0 indicating the lowest dissimilarity and 1 representing the perfect similarity. As dissimilarity indicates the presence of distortion, a critical factor in assessing image quality, we subtract the SSIM map from an all-one matrix $I$ to obtain the dissimilarity weights, where larger values signify more severe distortions. Finally, we calculate the Hadamard product between the distorted image and the dissimilarity weights to obtain the dissimilarity map. With this dissimilarity map as an input, the subsequent image quality evaluator can more effectively leverage the characteristics of IGM for BIQA. We call our method as distortion-based BIQA or D-BIQA.

---

**Algorithm 1:** Training DDPM

$\theta \leftarrow$ Randomly Initialize parameters of DDPM
$T, \beta_t \leftarrow$ Iinitialize time steps and variance schedule
**repeat**
    $(x_0, y) \leftarrow$ Get minibatch from a training dataset
    $t \leftarrow$ Uniform $(\{1,2,\ldots,T\})$
    $\epsilon \leftarrow \mathcal{N}(0, I)$
    $g \leftarrow \nabla \mathcal{L}$ Calculate gradient
    $\theta \leftarrow$ Update parameters
**until** Epoch is completed
**return** $\theta$

---

**Algorithm 2:** Inferencing with DDPM

$\theta \leftarrow$ Load parameters from the trained DDPM
$T, \beta_t \leftarrow$ Iinitialize time steps and the variation schedule
$x_T \sim \mathcal{N}(0, I)$
**for** $t = 1,2,\ldots,T$
    $z \sim \mathcal{N}(0, I)$ if $t > 1$ else $z = 0$
    $x_{t-1} = \frac{1}{\sqrt{\alpha_t}}\left(x_t - \frac{1-\alpha_t}{\sqrt{1-\bar{\alpha}_t}}D_\theta(x_t, y, t)\right) + \sigma_t z$
**end**
**return** Primary content

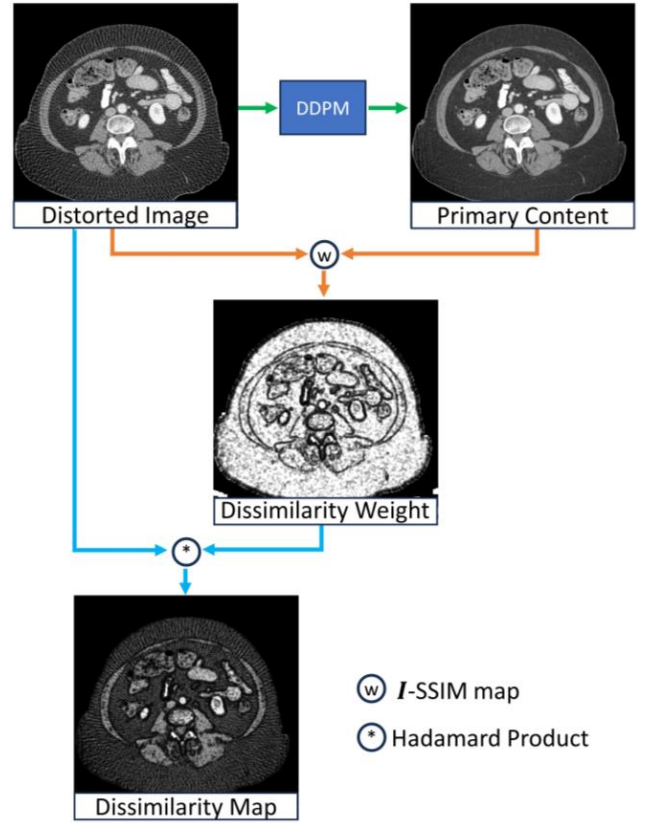

Figure 2. Dissimilarity map computed from the distorted image and the corresponding primary content.

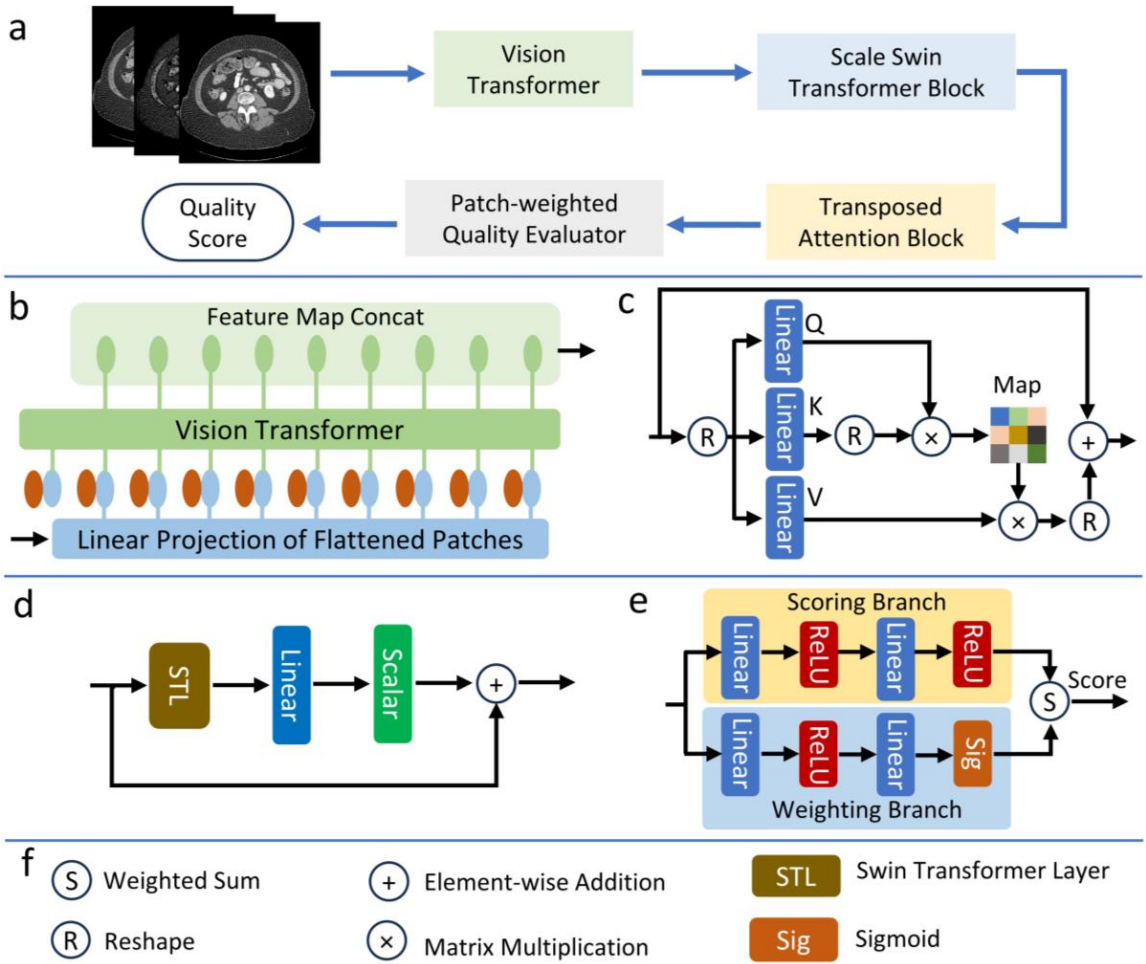

Figure 3. Transformer-based quality evaluator. a: general overview of the architecture; b: multi-scale vision transformer; c: transposed attention block; d: scale Swin transformer block; e: patch-weighted quality prediction; and f: legend of the submodules.

## B. Transform-based Quality Evaluator

After obtaining the dissimilarity map, we integrate it with the distorted image, resulting in a multi-channel image (a tensor in general). Our primary goal is to create a quality evaluator that can effectively process the multi-channel data tensor consisting of these input images. To comprehensively utilize information with both the spatial and channel dimensions, we introduce a transformer-based model tailored for perceptual quality prediction based on MANIQA [39]. The overall architecture of this transformer-based quality evaluator is depicted in Fig. 3a. The core components of the transformer-based quality evaluator include: 1) vision transformer; 2) transposed attention block; 3) scale Swin transformer block, and 4) patch-weighted quality prediction.

### 1) Vision Transformer

Fig. 3b shows the architecture of ViT. In the context of a given multi-channel image, denoted as $E \in \mathbb{R}^{H \times W \times 3}$, where $H$ and $W$ represent the height and width respectively, we employ the ViT denoted by $f_\varphi$ with a learnable vector of parameters $\varphi$. From the ViT, we extract features $F_i \in \mathbb{R}^{b \times c_i \times H_i W_i}$ corresponding to the i-th layer, where $i$ ranges from 1 to 12, $b$ denotes the batch size, $c_i$ represents channel size, $H_i$ and $W_i$ denote the dimensions of the i-th feature. Specifically, we utilize 4 layers to extract features with varying semantic significance. Then, we concatenate $\hat{F}_i$, where $i \in \{7,8,9,10\}$, yielding a composite feature tensor denoted as $\tilde{F} \in \mathbb{R}^{b \times \sum_i c_i \times H_i W_i}$.

### 2) Transposed Attention Block

To enhance channel interaction within these concatenated features, we employ a transposed attention block. Unlike the conventional spatial attention, this block facilitates self-attention across channels, effectively computing cross-covariance between channels to generate an attention map that encodes global contextual relations. Beginning with the feature tensor $\tilde{F} \in \mathbb{R}^{b \times \sum_i c_i \times H_i W_i}$, the transposed attention block first generates query (**Q**), key (**K**) and value (**V**) projections through independent linear projections. These projections are utilized to encode pixel-wise cross-channel dependency. The query and key projections are reshaped for dot-product interaction, producing a transposed-attention map $A \in \mathbb{R}^{\tilde{C} \times \tilde{C}}$. Note that $\tilde{C}$ is equal to $c_i$. It is noteworthy that the layer normalization and multi-layer perceptron components from the original transformer structure are omitted. Mathematically, the transposed attention block is defined as follows:

$$\widehat{X} = W_p Attn(\widehat{Q}, \widehat{K}, \widehat{V}) + X \tag{13}$$

$$Attn(\widehat{Q}, \widehat{K}, \widehat{V}) = \widehat{V} \cdot Softmax(\widehat{K} \cdot \widehat{Q}/\alpha) \tag{14}$$

where $\alpha$ denotes the spatial dimension of **Q**, **K** and **V**. The specifics of the transposed attention block are illustrated in Fig. 3c.

### 3) Scale Swin Transformer Block

The scale Swin transformer block is shown in Fig. 3d. The scale Swin transformer block consists of Swin transformer layers and convolutional layers. Given an input feature $F_{i,0}$, the scale Swin transformer block first encodes the feature tensor through 2 Swin transformer layers:

$$F_{i,j} = H_{STL_{i,j}}(F_{i,j}), j = 1,2, \tag{15}$$

where $H_{STL_{i,j}}(\cdot)$ represents the j-th Swin transformer layer in the i-th stage, $i \in \{1,2\}$. After the encoding process, a convolutional layer is applied before the residual connection. The output of the scale Swin transformer block is formulated as

$$F_{out} = \alpha \cdot H_{CONV}\left(H_{STL}(\tilde{F}_{i,2})\right) + \tilde{F}_{i,0} \tag{16}$$

where $H_{CONV}(\cdot)$ is the convolutional layer, and $\alpha$ denotes a scale factor for the output of the Swin transformer layer.

### 4) Patch-Weighted Quality Prediction

A dual-branch structure for patch-weighted quality prediction is shown in Fig. 3e. Given the feature tensor, we generate weight and score projections, which are achieved through 2 independent linear projections. The final patch score of the distorted image is generated by multiplying the score and weight of each patch, then the final score of the whole image is generated by the summation of all the final patch scores.

## C. Dataset

### 1) DDPM Training Dataset

For training DDPM, we simulated images with varying levels of dosage, by adding realistic noise to the dataset used in the 2016 low-dose CT grand challenge [56]. With the assumed use of a monochromatic source, the projection measurements from a CT scan follow the Poisson distribution, which can be expressed as

$$n_i \sim Poisson\{b_i e^{-l_i} + r_i\}, \quad i = 1, \cdots, I \tag{17}$$

where $n_i$ is the measurement along the i-th ray path. $b_i$ are the air scan photons, $r_i$ denotes read-out noise. In Eq. (17), the noise level can be controlled by $b_i$. This allowed us to obtain three additional noise levels equivalent to 50%, 25% and 10% of the normal dose for the 2016 low-dose CT grand challenge dataset. Streak artifacts are generated using a similar pipeline to noise insertion, but by reconstructing with different numbers of projections. There are three different numbers of projection views used in our study, which are 720, 360, and 180 views equiangularly distributed over a full scan. For each noise level we obtain 3 different sparse view, hence we obtained 12 reconstructed images for each normal dose scan. We selected 1,222 abdomen normal dose images from the 2016 low-dose CT grand challenge dataset and simulated 14,664 images with different noise levels and numbers of projection views to train our DDPM model.

### 2) Low-dose CT BIQA Dataset

The low-dose CT BIQA dataset comprises a total of 1,000 distorted abdominal images, which exhibit both noise and streak artifacts. These images were generated at four different dose levels: 100%, 50%, 25%, and 10%. Additionally, three numbers of projection views were uniformly selected over a full scan, resulting in a total of 12 different types of image degradation. Out of these images, 900 were allocated for the training phase, while the remaining 100 were reserved for the testing phase. In this dataset, CT values exceeding 350 Hounsfield Units (HU) were capped at 350 HU, resulting in a CT value range from -1000 HU to 350 HU. These CT values were then transformed to a normalized range from 0 to 1. The CT image quality assessment was performed using the abdominal soft-tissue window, defined by the width/level setting of 350/40, and evaluated by five proficient radiologists. The ultimate human perceptual score for each image was determined by averaging the individual ratings by these radiologists. The specific criteria for this assessment are detailed in Table I.

Table I. Image scoring criteria.

| Score | Quality | Diagnostic quality criteria |
|---|---|---|
| 0 | Bad | Desired features are not shown |
| 1 | Poor | Diagnostic interpretation is impossible |
| 2 | Fair | Suitable for compromised interpretation |
| 3 | Good | Good for diagnostic interpretation |
| 4 | Excellent | Anatomical features are clearly visible |

## D. Network Training

For training the DDPM model, we set the total number of time steps $T$ to 1,000. The model underwent training using the Adam optimizer, with a learning rate of $1 \times 10^{-4}$. The training process demonstrated convergence after $5 \times 10^5$ iterations on a computing server equipped with four Nvidia Tesla V100 GPUs. Upon completing the training process, we employed the trained DDPM to obtain the primary contents from the low-dose CT BIQA challenge dataset. Subsequently, we computed the dissimilarity map, resulting in a multi-channel image.

To train the transformer-based quality evaluator, we selected the ViT-B/8 model as our pre-trained model. This model was initially trained on ImageNet-21k and fine-tuned on ImageNet-1k, using a patch size of 8. To accommodate the varying input sizes of our datasets, we applied central cropping to resize the images to 448×448, as the edges of the images typically represent air and have limited influence on the overall image quality score. Furthermore, we resized these cropped images to a final size of 224×224. Then, we combined a dissimilarity map and two duplicated distorted images to form a three-channel image tensor.

The transformer-based quality evaluator consists of two stages, each comprising 2 transposed attention blocks and 1 scale Swin transformer block. The dimensions of the hidden layer, the number of heads, and the window size are set to 768, 4, and 4 in each scale Swin transformer block. We set the scale parameter to 0.8 in the scale Swin transformer block. For training, we established the learning rate at $1 \times 10^{-5}$ and used a batch size of 8. The Adam optimizer was used with a weight decay of $1 \times 10^{-5}$ and cosine annealing for learning rate scheduling. The selected loss function was the mean square error (MSE). As our transformer-based quality evaluator was built upon the foundation of MANIQA [39], we still call the transformer-based quality assessment method without primary content as MANIQA throughout this paper. Our experiments were conducted on an NVIDIA RTX A4000 using PyTorch 2.0.1 and CUDA 11.8 for both training and testing.

### E. Evaluation Criteria

We use Pearson's linear correlation coefficient (PLCC), the absolute value of the Spearman's rank order correlation coefficient (SROCC), and the Kendall rank-order correlation coefficient (KROCC) as the metrics to evaluate the performance of our models. The PLCC is defined as

$$PLCC = \frac{\sum_{i=1}^{M}(s_i - \mu_{s_i})(\hat{s}_i - \mu_{\hat{s}_i})}{\sqrt{\sum_{i=1}^{M}(s_i - \mu_{s_i})^2}\sqrt{\sum_{i=1}^{M}(\hat{s}_i - \mu_{\hat{s}_i})^2}} \quad (18)$$

where $s_i$ and $\hat{s}_i$ respectively indicate the ground-truth and predicted quality scores of the i-th image, $\mu_{s_i}$ and $\mu_{\hat{s}_i}$ indicate their means, and $M$ denotes the testing images. Let $d_i$ denote the difference between the ranks of the i-th test image in ground-truth and the predicted quality score. The SROCC is defined as

$$SROCC = 1 - \frac{6\sum_{i=1}^{M} d_i^2}{M(M^2 - 1)} \quad (19)$$

The KROCC is defined as:

$$KROCC = \frac{2(M_c - M_d)}{M(M - 1)} \quad (20)$$

where $M_c$ is the count of data pairs sharing the same rank, and $M_d$ is the count of data pairs with different rank. We use the absolute values of all the metrics, PLCC, SROCC and KROCC, are all in the range of [0, 1]. A higher value indicates better performance. The overall metric is defined as

$$Overall = PLCC + SROCC + KROCC \quad (21)$$

## III. RESULTS

### A. Inspection of Primary Contents

As obtaining high-quality primary contents is of paramount importance in mimicking the HVS, our initial focus is on evaluating primary contents across various quality scores and comparing the outcomes using distinct reconstruction methods. We employed the four classic methods for this purpose: 1) FBP: Reconstructing the original image from the low-dose CT BIQA challenge dataset using FBP; 2) RED-CNN: Utilizing CNN for low-dose CT reconstruction; 3) SU-Net: Combining the U-Net architecture with the Swin transformer for low-dose CT reconstruction; and 4) DDPM: Known for its powerful image generation capabilities in low-dose CT reconstruction.

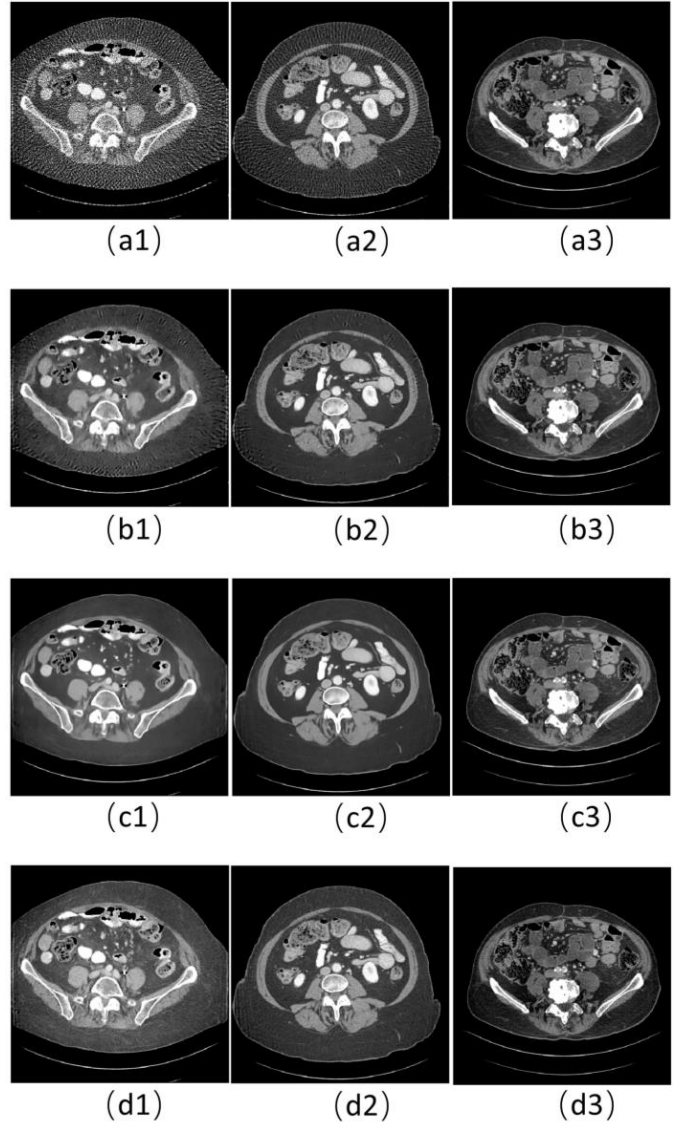

Figure 4. Visualization of the primary contents produced using different reconstructed methods. (a1)–(a3) the input distorted images at different distortion levels, of which quality scores are 0.2, 2.0 and 4.0, respectively; (b1)–(b3) the primary contents reconstructed by RED-CNN; (c1)–(c3) the primary contents reconstructed by SU-Net; and (d1)–(d3) the primary contents reconstructed by DDPM. The display window (width/level) is set to 350/40 HU.

In Fig. 4, we present three representative samples from the low-dose CT BIQA challenge dataset, with assigned quality scores of 0.2, 2.0, and 4.0 respectively. It can be seen that the FBP-reconstructed image with a quality score of 0.2 exhibits distortions caused by both noise and streak artifacts. Crucial features are not discernible. At a quality score of 2.0, the FBP-reconstructed image displays noticeable streak artifacts, but the main structural features are still recognizable. In the FBP reconstruction with a quality score of 4.0, the anatomical structure is highly visible, qualified as a reference image. The RED-CNN method successfully mitigates noise artifacts but sometimes amplifies streak artifacts. This effect is particularly pronounced in the image with a quality score of 0.2, exerting a negative impact on image quality. SU-Net effectively suppresses both noise and streak artifacts. However, the resultant image appears overly smooth, deviating from the authentic appearance of the real clinical counterpart. The DDPM approach emerges as an accurate yet robust solution, effectively suppressing noise and streak artifacts while simultaneously preserving intricate details. Furthermore, the distribution of the images closely aligns with that of real clinical reference images. As a result, opting for DDPM to generate high-quality primary contents proves feasible and advantageous, facilitating an accurate emulation of the IGM principle within the HVS.

### B. Visualization of the Prior Information

By employing primary contents generated through DDPM, we can glean insights into the proposed method. In Fig. 5, the first row showcases distorted images across varying levels of distortions. From the second to the last row, we present corresponding primary contents and dissimilarity maps. The effectiveness of the IGM becomes evident in its ability to mitigate noise and infer high-quality primary content, as evidenced in Fig. 5(d)-(f). Concurrently, as depicted in Fig. 5(g)-(i), when subjected to distinct levels of artifacts, the dissimilarity map exhibits diverse patterns. Notably, Fig. 5(i) demonstrates lower dissimilarity, implying less content degradation in Fig. 5(c).

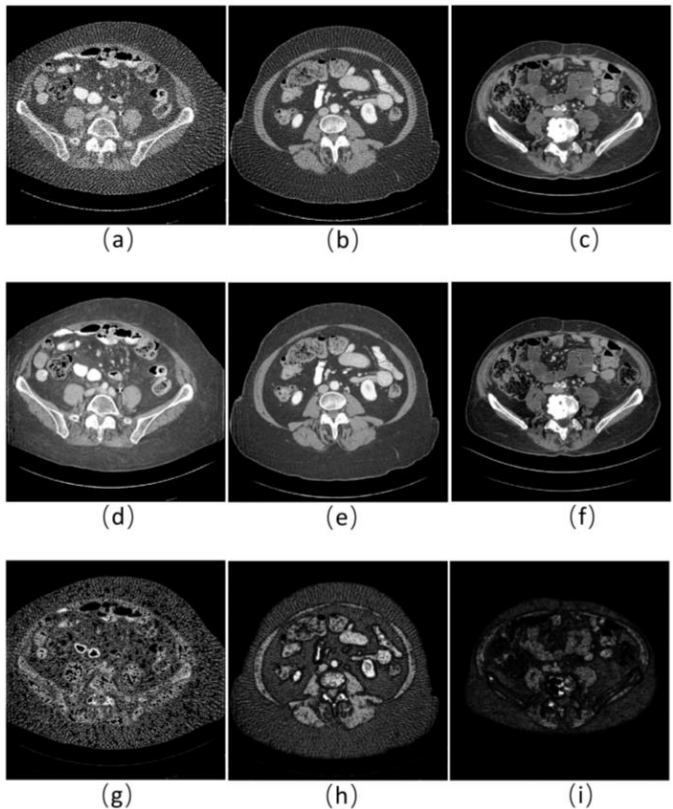

Figure 5. Generative prior information with DDPM. (a)–(c) the input distorted images at different distortion levels, with image quality scores of 0.2, 2.0 and 4.0 respectively. The lower the quality score, the higher the distortion level and the worse the perceptual quality; (d)–(f) the primary contents generated by DDPM; and (g)–(i) the dissimilarity maps calculated by SSIM.

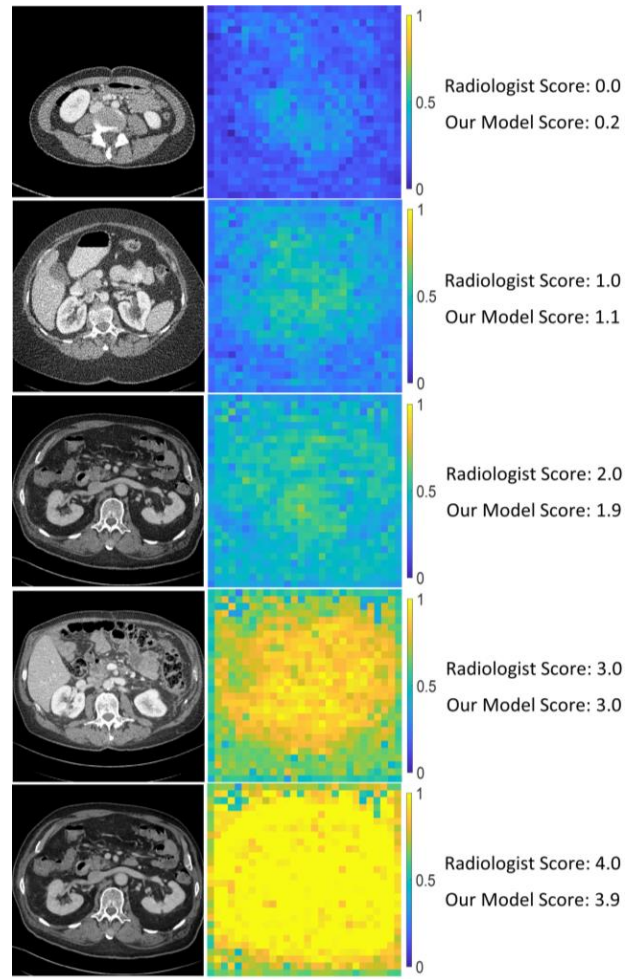

Figure 6. Visualization of several exemplary images from the test dataset. From left to right are the distorted images, the weight maps, and scores. From top to bottom are five representative example images with the radiologist scores of 0.0, 1.0, 2.0, 3.0 and 4.0 and our model predictions respectively.

### C. Visualization of Exemplary Images

In Fig. 6, we visualize distorted images and weight maps for five represetative examples from the low-dose CT BIQA test dataset. These weight maps clearly highlight salient subjects, which are significant areas that strongly influence human perception. When humans view an image, features of interest greatly affects our perception, and therefore these features receive higher weights as depicted in Fig. 6. Furthermore, as the score increases, the weight also becomes larger. Regarding the predicted scores for the images, we observed some bias relative to the radiologists' scores, indicating that there is still room for improvement.

### D. Performance Comparison within the Low-dose CT BIQA Challenge Dataset

Given the absence of reference images in the dataset, we straightforwardly partition the dataset based on distorted images. We compared our method against one traditional BIQA method and two CNN-based methods, along with one transformer-based method. For the traditional approach, we selected the natural image quality evaluator (NIQE) as a represetative method as implemented in the official code.

Among the CNN-based methods, we used DBCNN and Hyper-IQA as representative networks, using the implementations provided in [56]. As for the transformer-based method, we employed MANIQA. Referring to Table II, limited by the handcrafted features, the traditional NIQE demonstrates the lowest performance. The CNN-based methods show remarkable improvements over the NIQE approach. Between these two CNN-based methods, DBCNN outperforms in PLCC and KROCC metrics, while Hyper-IQA excels in SROCC. The performance of these two CNN-based methods is closely comparable. As a representative of transformer-based methods, MANIQA exhibits superior performance over the CNN-based methods. Remarkably, our proposed D-BIQA method achieves the highest performance. In summary, D-BIQA demonstrates strong efficacy on the low-dose CT BIQA challenge dataset, thereby affirming the effectiveness of our proposed approach.

Table II. Comparisons on the Low-dose CT BIQA Dataset.

| Methods | PLCC | SROCC | KROCC | Overall |
|---|---|---|---|---|
| NIQE | 0.9181 | 0.9335 | 0.7897 | 2.6414 |
| DBCNN | 0.9716 | 0.9693 | 0.8692 | 2.8103 |
| HyperIQA | 0.9680 | 0.9694 | 0.8672 | 2.8045 |
| MANIQA | 0.9789 | 0.9792 | 0.9041 | 2.8622 |
| D-BIQA | **0.9814** | **0.9816** | **0.9122** | **2.8753** |

## IV Discussions and Conclusion

Given the widespread artifacts in low-dose CT images, we have opted for a transformer-based approach to comprehend these artifacts globally. It is worth emphasizing the importance of pretraining when utilizing transformers. In this study, we have chosen to use a pretrained ViT-B/8 model with a 224×224 configuration for further training on a low-dose CT dataset. In typical natural image BIQA, researchers often employ a random crop of 224×224 to align with the requirement for training a transformer. However, in the case of CT images, the peripheral region of the image predominantly corresponds to air and should not influence the quality assessment. As a result, we have cropped each original $512 \times 512$ images down to a centralized $448 \times 448$ image, excluding the air and less important features from consideration. To ensure the quality of neural network training while simultaneously minimizing computational complexity, we have employed a common down-sampling technique [57]. Specifically, all images were consistently down-sampled to a uniform size of 224×224. This enables us to effectively address the globalized artifacts in low-dose CT images while losing little information in this medical imaging domain.

Directional artifacts in low-dose CT images can occasionally mislead radiologists because they might resemble actual anatomical structures. To address this issue, we have implemented a DDPM to mitigate these artifacts. While the DDPM has demonstrated the capability to generate high-quality images that closely resemble reality, it is important to clarify that our intention is not to restore a perfectly pristine or distortion-free image. Our DDPM efforts may still retain compromised information when dealing with severely distorted images. Nevertheless, our primary objective is to emulate the active inference process employed by IGM to predict the vital content within the image. As long as the DDPM results are helpful to derive the primary content and guide our image quality assessment, our goal is achieved.

In conclusion, we have introduced a novel D-BIQA model for image quality assessment based on DDPM-driven active inference. Leveraging the amazing image synthesis capability of DDPM, our active inference module seems adeptly emulating the IGM theory to forecast the primary contents from distorted images. Through the amalgamation of information derived from distorted images and dissimilarity maps, a multi-channel image tensor is formed and fed into a transformer-based quality evaluator. A comprehensive set of experiments substantiates the efficacy and superiority of our proposed approach. While we achieved the second place in the MICCAI 2023 low-dose CT perceptual image quality assessment grand challenge with our reported transformer-based quality evaluator, this much-improved approach for D-BIQA has further enhanced performance guided by the IGM theory, which is the main innovation of this work.

## Acknowledgments

The authors would like to express their gratitude to the organizers of the MICCAI 2023 low-dose CT perceptual image quality assessment grand challenge for providing the dataset.

## References

[1] H. Chen, Y. Zhang, M. K. Kalra, F. Lin, Y. Chen, P. Liao, J. Zhou, and G. Wang, "Low-dose CT with a residual encoder-decoder convolutional neural network," *IEEE Transactions on Medical Imaging*, vol. 36, no. 12, pp. 2524–2535, 2017.

[2] K. H. Jin, M. T. McCann, E. Froustey, and M. Unser, "Deep convolutional neural network for inverse problems in imaging," *IEEE Transactions on Image Processing*, vol. 26, no. 9, pp. 4509–4522, 2017.

[3] Q. Yang, P. Yan, Y. Zhang, H. Yu, Y. Shi, X. Mou, M. K. Kalra, Y. Zhang, L. Sun, and G. Wang, "Low-dose CT image denoising using a generative adversarial network with Wasserstein distance and perceptual loss," *IEEE Transactions on Medical Imaging*, vol. 37, no. 6, pp. 1348–1357, 2018.

[4] E. Kang, W. Chang, J. Yoo, and J. C. Ye, "Deep convolutional framelet denoising for low-dose CT via wavelet residual network," *IEEE Transactions on Medical Imaging*, vol. 37, no. 6, pp. 1358–1369, 2018.

[5] H. Shan, Y. Zhang, Q. Yang, U. Kruger, M. K. Kalra, L. Sun, W. Cong, and G. Wang, "3-d convolutional encoder-decoder network for low-dose CT via transfer learning from a 2-d trained network," *IEEE Transactions on Medical Imaging*, vol. 37, no. 6, pp. 1522–1534, 2018.

[6] Y. Han, J. C. Ye, "Framing U-Net via deep convolutional framelets: Application to sparse-view CT," *IEEE Transactions on Medical Imaging*, vol. 37, no. 6, pp. 1418-1429, 2018.

[7] Z. Zhang, X. Liang, X. Dong, Y. Xie, G. Cao, "A sparse-view CT reconstruction method based on combination of DenseNet and deconvolution," *IEEE Transactions on Medical Imaging*, vol. 37, no. 6, pp. 1407-1417, 2018.

[8] H. Zhang, B. Liu, H. Yu, B. Dong, "MetaInv-Net: Meta inversion network for sparse view CT image reconstruction," *IEEE Transactions on Medical Imaging*, vol. 40, no. 2, pp. 621-634, 2020.

[9] D. Hu, J. Liu, T. Lv, Q. Zhao, Y. Zhang, G Quan, et al., "Hybrid-domain neural network processing for sparse-view CT reconstruction," *IEEE Transactions on Radiation and Plasma Medical Sciences*, vol. 5, no. 1, pp. 88-98. 2020.

[10] M. Lee, H. Kim, H. J. Kim, "Sparse-view CT reconstruction based on multi-level wavelet convolution neural network," *Physica Medica*, vol. 80, pp. 352-362, 2020.

[11] Z. Wang, A. C. Bovik, H. R. Sheikh, E. P. Simoncelli, "Image quality assessment: from error visibility to structural similarity," *IEEE Transactions on Image Processing*, vol. 13, no. 4, pp. 600-612. 2004.

[12] V. Kamble, K. M. Bhurchandi, "No-reference image quality assessment algorithms: A survey," *Optik*, vol. 126, no. 11-12, pp. 1090-1097, 2015.

10 IEEE TRANSACTIONS ON MEDICAL IMAGING, VOL. xx, NO. x, 2020
[13] S. Xu, S. Jiang, W. Min, "No-reference/blind image quality assessment: a survey," *IETE Technical Review*, vol. 34, no. 3, pp. 223-245, 2017.
[14] A. J. Smola and B. Schölkopf, "A tutorial on support vector regression," *Statistics Computing*, vol. 14, no. 3, pp. 199–222, 2004.
[15] A. K. Moorthy and A. C. Bovik, "Blind image quality assessment: From natural scene statistics to perceptual quality," *IEEE Transactions on Image Processing*, vol. 20, no. 12 2011
[16] G. Wang, Z. Wang, K. Gu, L. Li, Z. Xia, L. Wu, "Blind quality metric of DIBR-synthesized images in the discrete wavelet transform domain," *IEEE Transactions on Image Processing*, vol. 11, no. 29 pp. 1802-1814, 2019.
[17] C. Deng, S. Wang, A. C. Bovik, G. B. Huang, B. Zhao, "Blind noisy image quality assessment using sub-band kurtosis," *IEEE Transactions on Cybernetics*, vol. 50, no. 3, pp. 1146-1156, 2019.
[18] M. A. Saad, A. C. Bovik, C. Charrier, "Blind image quality assessment: A natural scene statistics approach in the DCT domain," *IEEE Transactions on Image Processing*, vol. 21, no. 8, pp. 3339-3352, 2012.
[19] A. Mittal, A. K. Moorthy, and A. C. Bovik, "No-reference image quality assessment in the spatial domain," *IEEE Transactions on Image Processing*, vol. 21, no. 12, pp. 4695-4708, Dec. 2012.
[20] A. Mittal, R. Soundararajan, and A. C. Bovik, "Making a 'completely blind' image quality analyzer," *IEEE Signal Processing Letter*, vol. 22, no. 3, pp. 209–212, 2013.
[21] M. Khan, I. F. Nizami, and M. Majid, "No-reference image quality assessment using gradient magnitude and wiener filtered wavelet features," *Multimedia Tools and Applications*, vol. 78, pp. 14485-14509, 2019.
[22] Q. Li, W. Lin, J. Xu, and Y. Fang, "Blind image quality assessment using statistical structural and luminance features," *IEEE Transactions on Multimedia*, vol. 18, no. 12, pp. 2457–2469, 2016.
[23] Q. Wu, Z. Wang, H. Li, "A highly efficient method for blind image quality assessment," *Proceedings of the IEEE International Conference on Image Processing*, pp. 339-343, 2015.
[24] M. Zhang, C. Muramatsu, X. Zhou, T. Hara, H. Fujita, "Blind image quality assessment using the joint statistics of generalized local binary pattern," *IEEE Signal Processing Letters*, vol. 22, no. 2, pp. 207-210, 2014.
[25] W. Xue, X. Mou, L. Zhang, A. C. Bovik, X. Feng, "Blind image quality assessment using joint statistics of gradient magnitude and Laplacian features," *IEEE Transactions on Image Processing*, vol. 23, no. 11, pp. 4850-4862, 2014.
[26] L. Kang, P. Ye, Y. Li, D. Doermann, "Convolutional Neural Networks for No-Reference Image Quality Assessment," *Proceedings of the IEEE Conference on Computer Vision and Pattern Recognition*, pp. 1733-1740, 2014.
[27] W. Hou, X. Gao, D. Tao, X. Li, "Blind image quality assessment via deep learning," *IEEE Transactions on Neural Networks and Learning Systems*, vol. 26, no. 6, pp. 1275-1286, 2014.
[28] X. Yang, F. Li, H. Liu, "A survey of DNN methods for blind image quality assessment," *IEEE Access*, vol. 7, pp. 123788-123806, 2019.
[29] X. Liu, J. Weijer, A. D. Bagdanov, "Rank-IQA: Learning from Rankings for No-Reference Image Quality Assessment," *Proceedings of the IEEE International Conference on Computer Vision*, pp. 1040-1049, 2017.
[30] W. Zhang, K. Ma, J. Yan, D. Deng, Z. Wang, "Blind image quality assessment using a deep bilinear convolutional neural network," *IEEE Transactions on Circuits and Systems for Video Technology*, vol. 30, no. 1, pp. 36–47, 2018.
[31] S. Su, Q. Yan, Y. Zhu, C. Zhang, X. Ge, J. Sun et al., "Blindly assess image quality in the wild guided by a self-adaptive hyper network," *Proceedings of the IEEE Conference on Computer Vision and Pattern Recognition*, pp. 3667-3676, 2020.
[32] H. Zhu, L. Li, J. Wu, W. Dong, G. Shi, "Meta-IQA: Deep meta-learning for no-reference image quality assessment," *Proceedings of the IEEE Conference on Computer Vision and Pattern Recognition*, pp. 14143-14152, 2020.
[33] A. Dosovitskiy, L. Beyer, A. Kolesnikov, D. Weissenborn, X. Zhai, T. Unterthiner, "An image is worth 16x16 words: Transformers for image recognition at scale," *arXiv reprint* arXiv:2010.11929, 2020.
[34] J. You, J. Korhonen, "Transformer for image quality assessment," *Proceeding of the IEEE International Conference on Image Processing*, pp. 1389-1393, 2021.
[35] J. Ke, Q. Wang, Y. Wang, P. Milanfar, F. Yang, "MUSIQ: Multi-scale image quality transformer," *Proceedings of the IEEE International Conference on Computer Vision*, pp. 5148-5157, 2021.
[36] S. A. Golestaneh, S. Dadsetan, K. M. Kitani, "No-reference image quality assessment via transformers, relative ranking, and self-consistency," *Proceedings of the IEEE Winter Conference on Applications of Computer Vision*, pp. 1220-1230, 2022.
[37] M. V. Conde, M. Burchi, R. Timofte, "Conformer and blind noisy students for improved image quality assessment," *Proceedings of the IEEE Conference on Computer Vision and Pattern Recognition*, pp. 940-950, 2022.
[38] J. Wang, H. Fan, X. Hou, Y. Xu, T. Li, X. Lu, "MSTRIQ: No reference image quality assessment based on Swin transformer with multi-stage fusion," *Proceedings of the IEEE Conference on Computer Vision and Pattern Recognition*, pp. 1269-1278, 2022.
[39] S. Yang, T. Wu, S. Shi, S. Lao, Y. Gong, M. Cao, et al., "MANIQA: Multi-dimension attention network for no-reference image quality assessment," *Proceedings of the IEEE Conference on Computer Vision and Pattern Recognition*, pp. 1191-1200, 2022.
[40] L. Yu, J. Li, F. Pakdaman, M. Ling, M. Gabbouj, "MAMIQA: No-Reference Image Quality Assessment Based on Multiscale Attention Mechanism with Natural Scene Statistics," *IEEE Signal Processing Letters*, vol. 30, pp. 588-592, 2023.
[41] J. Gu, H. Cai, C. Dong, J. S. Ren, R. Timofte, Y. Gong et al., "NTIRE 2022 challenge on perceptual image quality assessment," *Proceedings of the IEEE Conference on Computer Vision and Pattern Recognition*, pp. 951-967, 2022.
[42] G. Zhai, X. Wu, X. Yang, W. Lin, and W. Zhang, "A psychovisual quality metric in free-energy principle," *IEEE Transactions on Image Processing*, vol. 21, no. 1, pp. 41–52, Jan. 2012.
[43] J. Wu, W. Lin, G. Shi, and A. Liu, "Perceptual quality metric with internal generative mechanism," *IEEE Transactions on Image Processing*, vol. 22, no. 1, pp. 43–54, Jan. 2013.
[44] K. Gu, G. Zhai, X. Yang, and W. Zhang, "Using free energy principle for blind image quality assessment," *IEEE Transactions on Multimedia*, vol. 17, no. 1, pp. 50–63, Jan. 2015.
[45] J. Ma, J. Wu, L. Li, W. Dong, X. Xie, G. Shi, "Blind image quality assessment with active inference," *IEEE Transactions on Image Processing*," vol. 30, pp. 3650-3663, 2021.
[46] M. P. Eckert and A. P. Bradley, "Perceptual quality metrics applied to still image compression," Signal Processing, vol. 70, no. 3, pp. 177–200, 1998.
[47] J. Ho, A. Jain, P. Abbeel, "Denoising diffusion probabilistic models," *Advances in Neural Information Processing Systems*, vol. 33, pp. 6840-51, 2020.
[48] C. Saharia, J. Ho, W. Chan, T. Salimans, D. J. Fleet, M. Norouzi, "Image super-resolution via iterative refinement," *IEEE Transactions on Pattern Analysis and Machine Intelligence*, vol. 45, no. 4, pp. 4713-4726, 2022.
[49] P. Dhariwal, A. Nichol, "Diffusion models beat GANs on image synthesis," *Advances in Neural Information Processing Systems*, vol. 34, pp. 8780-94, 2021.
[50] W. Xia, Q. Lyu, G. Wang, "Low-Dose CT Using Denoising Diffusion Probabilistic Model for 20x Speedup," *arXiv preprint* arXiv:2209.15136, 2022.
[51] Y. Shi, G. Wang, "Conversion of the Mayo LDCT Data to Synthetic Equivalent through the Diffusion Model for Training Denoising Networks with a Theoretically Perfect Privacy," *arXiv preprint* arXiv:2301.06604, 2023.
[52] Q. Gao, S. Li, M. Zhu, D. Li, Z. Bian, Q. Lyu, et al., "Blind CT image quality assessment via deep learning framework," *Proceeding of the IEEE Nuclear Science Symposium and Medical Imaging Conference*, pp. 1-4, 2019.
[53] S. Li, J. He, Y. Wang, Y. Liao, D. Zeng, Z. Bian, et al., "Blind CT image quality assessment via deep learning strategy: initial study," *Proceeding of Society of Photo-Optical Instrumentation Engineers*, vol. 10577, pp. 293-297, 2018.
[54] Online available: https://ldctiqac2023.grand-challenge.org/.
[55] O. Ronneberger, P. Fischer, T. Brox, "U-net: Convolutional networks for biomedical image segmentation," *In International Conference on Medical Image Computing and Computer-Assisted Intervention*, vol. 9351, pp. 234-241, 2015.
[56] C. Chen, Pytorch toolbox for image quality assessment, 6, 2022.
[57] H. R. Sheikh, A. C. Bovik, G. D. Veciana, "An information fidelity criterion for image quality assessment using natural scene statistics," *IEEE Transactions on Image Processing*, vol. 14, no. 12, pp. 2117-2128, 2005.